\documentclass[12pt]{article}

\usepackage[margin=1in]{geometry}
\usepackage[T1]{fontenc}
\usepackage[utf8]{inputenc}
\usepackage{graphicx}
\usepackage{url}

\begin{document}

\title{AI-augmented science and the new institutional scarcities}

\author{Lauri Lov\'en\\[2pt]
\normalsize Future Computing Group, University of Oulu, Oulu, Finland\\
\normalsize lauri.loven@oulu.fi \quad ORCID: 0000-0001-9475-4839}
\date{}

\maketitle

\begin{abstract}
\noindent Artificial intelligence now produces convincing-looking scientific judgment (reviews, rankings, attributions, verifications) at almost no marginal cost. An influential reading of AI economics holds that prediction becomes cheap while human judgment stays scarce. For science, that reading understates the problem: what has become cheap is a counterfeit of judgment itself. This matters most for institutions whose product is trusted judgment, which is what journals, universities, funders, and learned societies exist to manufacture. They do not merely adapt to the technology; they compete with it for the same functional role. Four things become scarce instead: verified signal, legitimacy, authentic provenance, and integration capacity. Integration capacity means how much AI-delegated judgment a scientific community will accept before it stops trusting the journals, panels, and conferences that admitted it. It is the least developed of the four and the most binding: better tooling cannot buy it. The way forward for AI-augmented science is not acceleration but the redesign of its certifying infrastructure around these new scarcities.
\end{abstract}

\medskip
\noindent\textbf{Keywords:} AI-augmented science, scientific institutions, sociology of science, peer review, reproducibility, content provenance, commons governance, credentialing

\section{The scarcity-structure frame}\label{sec:frame}

When a frontier AI model triages a manuscript, drafts a peer review, scores a grant proposal, or checks whether a numerical claim reproduces, it produces competent-looking judgment at scale, at marginal cost approaching zero for its textual outputs. This is no longer hypothetical: peer-reviewed multi-agent systems now generate and rank novel research hypotheses, and design and interpret wet-lab experiments with minimal human intervention~\cite{ghareeb2026robin,aygun2026era}. What such systems produce is not yet reliable judgment. It is output that is hard to distinguish from judgment at the surface of the text, and that distinction is the hinge of this Comment.

Every technological revolution can be read through one diagnostic: what moved from scarce to abundant, and which institutions reorganised around the shift~\cite{perez2002}. Earlier revolutions made mechanical power, then signal, abundant, and rebuilt institutions around each. The received one-liner is that technology moves fast while institutions do not keep up. This Comment treats that not as a law but as a hypothesis to be examined for AI, and the examination turns on which scarcity the technology actually inverts.

In the influential reading that Agrawal, Gans, and Goldfarb popularised~\cite{agrawal2022}, AI collapses the cost of \textit{prediction}, while \textit{judgment} (in plain terms, knowing what to do with a prediction) remains the expensive human complement. The argument holds for narrow predictive systems sold into routine decision pipelines. For scientific institutions it needs sharpening rather than rejection. In that framework's own terms, a model that drafts a review is producing a prediction of what a competent judge would conclude. What has collapsed in cost is therefore not judgment but its counterfeit: an output that mimics the scarce complement without carrying its verification. When cheap counterfeits of a certified good circulate, certification stops separating the genuine article from the imitation, and the value of the certificate erodes. The scarce complements are then no longer judgment alone but the four goods developed below.

Land, mechanical power, and signal are inputs that institutions coordinate; judgment is what scientific institutions \textit{manufacture}. Universities certify scholarly judgment. Journals certify the validity of claims. Granting agencies certify which proposals deserve funding, and learned societies certify professional competence. When competent-looking judgment becomes abundant, these institutions cannot re-allocate their product elsewhere the way a factory re-purposes land or power. Their only fall-back is to change what they certify: from judgment to verified, legitimate judgment. That is institutional redesign rather than market re-allocation, and it depends on shared infrastructure no single institution can build alone. Nor is displacement automatic: the technology competes with the certifying institutions only insofar as audiences come to credit machine outputs as judgment, which is why the fourth scarcity below, integration capacity, is the binding one. AI augmentation thus does not merely accelerate science; it changes what science \textit{is for} as an institutional form. A general, cross-domain version of the scarcity taxonomy is developed elsewhere~\cite{loven2026cacm}; the argument of this Comment (the mapping from each certifying institution of science to the scarcity assumption it encodes, the per-institution diagnosis, and the verification-first agenda) appears here for the first time and stands on its own.

\section{What was scarce, in science, until now}\label{sec:was}

Whatever their heterogeneous historical origins, the core institutions of modern science currently function as devices for managing one scarcity: qualified human attention. The historical record cautions against any stronger design claim; refereeing, for instance, emerged late, unevenly, and for reasons that had as much to do with the public standing of learned societies as with triage~\cite{csiszar2018}. But the functional reading holds today, and it is the current function, not the origin, that AI destabilises. Peer review triages claims to a small number of qualified readers before the rest of the community spends time on them. Authorship norms attribute cognitive contributions to identifiable individuals, so that credit and responsibility travel together. Reproducibility practice rests on the assumption that publication is a probabilistic signal of truth, because direct verification is expensive and the community cannot afford to verify everything. Grant evaluation allocates scarce funding through equally scarce panel capacity, using triage heuristics that presume both proposers and reviewers are cognitively bounded. The same structural pressure shows up across the credentialed professions more broadly~\cite{susskind2022}; science is one profession among several exposed to this shift, with the further wrinkle that its credentialing body is the literature itself.

Each institution encodes a scarcity assumption. When the assumption holds, the institution produces legitimate signal and scales gracefully. When it fails, the institution degrades from the inside, often before anyone notices, because surface activity (submissions, reviews, citations, metrics such as h-indices) continues unabated while the underlying function hollows out. The machine-learning reproducibility-programme literature~\cite{pineau2021reproducibility} is in part a record of exactly this slow-motion hollowing, before AI arrived to amplify it. AI augmentation does not introduce the failure mode; it forces the institutional question that the reproducibility debate had already raised.

\section{What changes when competent-looking judgment is abundant}\label{sec:changes}

AI augmentation does not remove the scarcity of attention. It removes the scarcity of competent-looking judgment applied to that attention. Much of the strain predates AI: submission volumes at the major machine-learning conferences (NeurIPS, ICML, ICLR) grew several-fold over the past decade, and controlled experiments found reviewer verdicts substantially inconsistent long before language models entered the loop~\cite{cortes2021}. What AI adds is an amplifier, collapsing the marginal cost of competent-looking text on both sides of the review transaction. The failure modes appear first in the AI/ML community's own venues, which makes that community a leading indicator for disciplines with slower publication cycles.

\textit{Peer review} faces a volume problem that does not converge. Submission volume scales with the availability of AI-assisted writing, and AI-assisted review scales in parallel: an estimated 6.5 to 16.9 percent of reviews at recent AI conferences were already substantially machine-modified~\cite{liang2024}. Review quality degrades unless the process is redesigned. The queue becomes an arms race between machine-assisted writing and machine-assisted screening, in which neither side gains durable ground at the rhetorical layer, and the signal at the end is worth less than the signal that went in.

\textit{Authorship} faces an attribution problem that disclosure does not solve. Team science broke the named-authors-on-the-byline abstraction long before AI, and contributorship taxonomies such as CRediT~\cite{brand2015} are the existing institutional response; but those taxonomies presume that every contributor is a person who can bear responsibility. The human-in-the-loop model fails when the intellectual substance is co-produced across a human-AI workflow neither party fully audits~\cite{birhane2023science}. Naming the language model is necessary but does not resolve who contributed what, and no contributor role can carry responsibility for output nobody verified.

\textit{Reproducibility} faces an economic inversion: the cost of producing plausible-looking, unreproducible work is falling faster than the cost of verifying it. Publication then risks ceasing to be a reliable probabilistic signal of truth. AI/ML benchmarks absorbed this hit early, through test-set contamination (evaluation data leaking into training corpora), leaderboard climbing tuned to public rankings rather than to the underlying capability, and irreproducible state-of-the-art claims~\cite{kapoor2023leakage}.

\textit{Grant evaluation} faces the same volume-versus-depth pathology with an added twist: AI-assisted proposals converge on the rhetorical surface of fundability (the right vocabulary, the right structure, the right citations) without necessarily matching in intellectual depth. The AI funding programmes themselves (NSF AI Institutes, EU Horizon Europe AI calls, UK ARIA) are exposed first: their proposal templates and panel-reading protocols were not designed for an environment where rhetorical surface and intellectual depth diverge.

Each is an institutional failure mode that compounds silently, not a productivity complaint. Treating any of them as a tooling problem is a category error: the binding constraint is institutional.

\section{The new scarcities}\label{sec:new}

When AI-generated claims are abundant, four things become scarce and load-bearing, each mapping to a design problem current institutional forms do not solve.

\textit{Verified signal} is the first. A paper that has been reproduced (the computational claim re-run, the data re-checked, the numerical results re-derived) carries information that a paper merely ``reviewed'' does not. Verification infrastructure (reproducibility pipelines, provenance chains, attestation, adversarial replay) is the editorial work that actually adds information. Two boundary conditions keep the claim honest. First, what counts as a successful replication is itself often contested, the predicament known as the experimenter's regress~\cite{collins1985}: verification is strongest for closed computational claims and degrades as claims depend on tacit skill or interpretation, which is one reason certification must be tiered rather than binary (Section~5). Second, certification marks decouple from actual verification unless someone audits the verifiers: when one journal issue's open-data-badged articles were computationally checked, only one in fourteen fully reproduced~\cite{cruwell2023}. The reproducibility checklists deployed at NeurIPS and elsewhere~\cite{pineau2021reproducibility} are a partial version of the needed infrastructure; they were built when the marginal cost of a fabricated baseline was higher than it now is.

\textit{Legitimacy} is the second. By legitimacy this Comment means an audience-attributed asset of the institution: the generalised perception, among those who rely on its certificates, that its certifying acts are proper and its marks mean what they claim~\cite{suchman1995}. Merton's ``organized skepticism''~\cite{merton1973} matters here not as a description of how scientists behave but as science's professed ideology of self-correction, whose enactment (somebody actually checking) is what replenishes that perception. Legitimacy so understood behaves like a commons~\cite{ostrom1990,frischmann2014}, with a twist. It is not used up by readers (one reader's trust does not subtract from another's); it is polluted by bad certification, each visibly unreproducible accepted paper degrading the shared stock of trust in the certifying mark. Ostrom's design principles therefore transfer through their provision side, the rules organising who contributes maintenance labour, rather than their appropriation side~\cite{ostrom1990}: the free-rider problem is under-provision of checking, and the pollution problem is certification without it. The credit-rating agencies supply the cautionary case, stated precisely. Issuer-pays certification, in which the rated firm pays the rater much as an article-fee-paying author pays a journal, coexisted with informative corporate ratings for three decades, disciplined by the agencies' reputational stake~\cite{white2010}. The structured-finance collapse required activating conditions: products complex enough to permit shopping among raters, boom-time fee revenue that outweighed reputational caution, and, critically, regulation that hardwired demand for ratings into capital rules, making that demand insensitive to accuracy~\cite{white2010}. Scientific publishing already shares the payment structure of the cautionary case (gold open access is issuer-pays, and predatory journals are its ratings-inflation equilibrium); what it does not yet have is the hardwired demand, a point Section~5 returns to. Legitimacy-maintenance is the scarce labour, not certification itself. Institutions that overspend their legitimacy budget do not announce it; they become irrelevant while surface activity continues.

\textit{Authentic provenance} is the third. Which model produced a claim, which data it was trained on, which prompt sequence elicited it, which human decisions shaped the output: these are facts about a scientific artefact that current publication metadata does not carry. The components exist piecemeal: model cards and datasheets (standardised disclosure documents describing what a model or a dataset contains and how it was built) and W3C-PROV (a web standard for machine-readable provenance records). What is missing is a unified scientific-artefact standard that composes them. C2PA, an industry standard that cryptographically records a media file's origin and edit history, demonstrates the approach for images and video~\cite{c2pa}; extending it to multi-component scientific outputs spanning code, data, trained models, prompt records, and human-edited text is an open infrastructure problem for scholarly communication as a whole, and the AI/ML community has already built most of the components.

\textit{Integration capacity} is the fourth and least obvious scarcity. It is the ceiling on how much delegated AI-judgment a scientific community will absorb before readers stop trusting the journals, conferences, and grant panels that admitted it. The constraint is multi-stakeholder: authors, reviewers, readers, and downstream users of scientific claims (clinicians, policymakers, journalists, industry) each have a distinct tolerance curve, and the institution is bounded by the strictest of them. Whether these tolerances are being measured is itself an open question: candidate proxies (dissent rates, uptake of human-only review tracks where offered, disclosure-weighted citation patterns) are, to the author's knowledge, not yet systematically collected anywhere, and instrumenting them is a concrete assignment for the coalition proposed in Section~5. Three established constructs sit nearby, and the differences locate the claim. Absorptive capacity~\cite{cohenlevinthal1990} is a producer-side ability, an organisation's capacity to recognise, assimilate, and exploit external knowledge, grown by investment; integration capacity is audience-side, belonging to the community that consumes certification. Power's audit society~\cite{power1997audit} describes the saturating supply of verification rituals; integration capacity is the demand side, the willingness to accept what checking certifies. The closest historical account is Porter's study of mechanical objectivity~\cite{porter1995}, in which communities delegated judgment to impersonal procedures precisely where trust in experts was weak, and tolerance for the delegation grew as the procedures became infrastructural. The AI case differs in the decisive respect: Porter's procedures were standardised and inspectable, so every delegation carried its own verification and added to a public track record. Delegation to opaque, counterfeit-capable systems carries no such warrant. The conditioned claim is therefore that integration capacity is spent by unverified delegation and grown by delegation coupled to verification, which is exactly what the agenda of Section~5 is for. Integration capacity is also distinct from legitimacy: legitimacy is the institution's stock of accepted authority, while integration capacity is the audience-side rate limit on admitting delegated cognition before that stock drains. This is the scarcity no amount of better tooling can buy.

The four are consistent with the knowledge-commons decomposition (resource, community, governance, goals) of Frischmann, Madison, and Strandburg~\cite{frischmann2014}, with integration capacity as the addition prior frameworks do not name. Different domains exhibit the four in different mixes; clinical trials (registration, and the CONSORT reporting standard) and economics journals (mandatory replication packages) present sharper-resolved versions of the same structure.

\section{What a verification-first agenda would change}\label{sec:agenda}

Four moves follow concretely; none requires technology that does not already exist, and each is stated with its nearest existing instrument and the difference from it. Together they rebuild verified signal, provenance, and legitimacy. The AI/ML venues where the strain is furthest advanced are the natural site of first deployment, but the pattern generalises: for fields without AI/ML's tooling base, the entry move is the cheapest tier below (automated checking of numerical and citation content), adopted through the same editorial channels.

\textit{Reproducibility-first editorial action.} A subset of accepted papers carry a statement of the form ``we reproduced the computational claim,'' not merely ``we opined on the narrative.'' The nearest instruments are the ML Reproducibility Challenge and the artefact-evaluation tracks at systems conferences, where reviewers re-run submitted code; the delta is editorial weight and scale. The honest lesson of their track record is that evaluator labour, not technical capability, is the binding constraint, which is why financing appears below as a design requirement rather than a footnote. The recruitment mechanism is individual: investigators invest effort where it converts into recognition, grants, and standing (the credibility cycle~\cite{latour1986}), so verification labour must itself become creditable in hiring and research assessment, or the evaluator supply never materialises.

\textit{Tiered certification.} Separate claim-level, result-level, and framework-level certification, and publish the tier with the paper. AI-assisted tooling can largely handle claim-level checks (numerical agreement, citation integrity, code-runs-as-described); humans certify results and frameworks (whether the claim is interesting, whether the framing is sound, whether the contribution is novel). The commoditisation is the point: claim-level checks verify output against ground truth, so automation durably helps the checker there, while at the rhetorical layer the same technology helps writer and screener equally, in the arms race of Section~3. The nearest instruments are registered reports~\cite{chambers2022}, which certify a framework before results exist, and eLife's publish-review-curate model~\cite{eisen2022}, which publishes editorial assessments alongside papers; the delta is the verification tier, a published statement that claims were \textit{checked}, not merely evaluated. Their measured uptake, steady but slow across a decade, is the realistic adoption curve, and it is why the mandate lever below exists. The natural coordinator is a coalition of programme committees and editorial boards rather than any single journal, on the model of the medical journal editors' committee (ICMJE), whose 2004 registration requirement remade clinical-trial reporting~\cite{deangelis2004}.

\textit{Provenance attestation by default.} Every AI-augmented paper carries a machine-readable provenance record: model version, data sources, prompt record, human decision points, and the audit log connecting them. The components exist (Section~4); what is missing is the institutional commitment, plus a custodianship layer willing to own storage, signature verification, dispute resolution, and deprecation policy. Candidate custodians (arXiv, OpenReview, or a community-governed commons) are today resourced for none of these functions, so custodianship implies new money and a governance mandate, not a volunteer add-on. Regulation is a tailwind here rather than a competitor: the EU AI Act governs AI-system transparency, including machine-readable marking of synthetic content (Article 50), not the science-certification function this Comment addresses, and its marking obligations align with exactly this attestation infrastructure~\cite{euaiact2024}.

\textit{Verification as a commons.} Certification markets concentrate on their own: reputation is a slowly built asset, users coordinate on common standards, and a monopoly certifier profits by revealing as little as possible~\cite{lizzeri1999}. Verification labour pooled across journals and conferences resists that concentration, but only if designed to. Three design features carry the weight. First, governance is polycentric: many overlapping accreditors rather than one, with the communities subject to the rules holding seats in writing them, including resource-constrained and Global-South venues, per Ostrom's participation principle~\cite{ostrom1990}. Second, payment does not flow from the certified to the certifier: funder-financed verification services and revenue-scaled consortial dues, rather than author-side fees, so the commons does not inherit the issuer-pays conflict of Section~4. Third, mandated demand stays accuracy-sensitive. Funders (NSF, Horizon Europe, ARIA) can condition grant eligibility on certified-tier publication, the enforcement lever a learned-society convener otherwise lacks; but the credit-rating episode shows that hardwired demand is exactly how certification discipline dies, so the mandate must attach only to accreditors whose own verification accuracy is periodically audited, with every solicited certification disclosed to block shopping among accreditors. The realistic sequence follows the trial-registration precedent: a venue-level prototype tier first, then a coalition standard, then the funder condition~\cite{deangelis2004}.

These costs fall asymmetrically. Reproducibility infrastructure, provenance engineering, and tiered editorial labour are paid disproportionately by under-resourced institutions, by the Global South, and by disciplines without AI/ML's tooling base; the AI/ML community's leading-indicator position is also an incumbency advantage. Shared infrastructure as a redistributive commons, with explicit fee waivers and governance seats for resource-constrained venues, is the form the design must take to avoid contracting access while upgrading certification.

Partial implementations are not neutral: a journal that automates review without rebuilding verification commons accelerates the arms race rather than damping it (the general form of this composition argument is developed in~\cite{loven2026cacm}). None of these moves is cheap; all attack the binding scarcity rather than the decorative one. The journals and conferences that survive the next decade will not be those that automated their review queues fastest, but those that rebuilt their verification commons first. Producing competent-looking judgment is now the cheapest part of science; producing legitimate judgment is becoming the most expensive. The AI/ML community, as both producer of the abundance and its first institutional casualty, will run the first experiments. But the redesign is an institutional problem before it is a technical one, and the fields that study how institutions gain, spend, and lose legitimacy (the sociology of science, institutional economics, and the governance of commons) hold the design knowledge the redesign will need.

\section*{Data availability}
Data sharing is not applicable to this article, as no datasets were generated or analysed.

\section*{Author contributions}
The sole author conceived and wrote this Comment.

\section*{Competing interests}
The author declares no competing interests.

\section*{Ethical approval}
Not applicable: this article does not contain any studies with human participants or animals.

\section*{Informed consent}
Not applicable: this article does not contain any studies with human participants.

\bibliographystyle{unsrt}
\bibliography{references}

\end{document}